\documentclass[prd,nofootinbib,superscriptaddress]{revtex4-1}

\usepackage{amsmath, amssymb, amsthm, graphicx, epsfig, fancyhdr,epsfig, slashed, mathrsfs}

\usepackage{float}

\usepackage{mathtools} 

\usepackage{tikz,xcolor}
\usepackage[breaklinks,pdfpagelabels]{hyperref}
\hypersetup{colorlinks=true,linkcolor=blue,urlcolor=blue,citecolor=blue}

\definecolor{lime}{HTML}{A6CE39}
\DeclareRobustCommand{\orcidicon}{
	\begin{tikzpicture}
	\draw[lime, fill=lime] (0,0) 
	circle [radius=0.2] 
	node[white] {{\fontfamily{qag}\selectfont \tiny ID}};
	\draw[white, fill=white] (-0.0625,0.095) 
	circle [radius=0.007];
	\end{tikzpicture}
	\hspace{-2mm}
}

\foreach \x in {A, ..., Z}{\expandafter\xdef\csname orcid\x\endcsname{\noexpand\href{https://orcid.org/\csname orcidauthor\x\endcsname}
			{\noexpand\orcidicon}}
}

\newcommand{\be}{\begin{equation}}
\newcommand{\ee}{\end{equation}}
\newcommand{\bea}{\begin{eqnarray}}
\newcommand{\eea}{\end{eqnarray}}

\def\cob{\color{blue}}
\newcommand{\au}[2]{#1.~#2}
\newcommand{\arX}[1]{\href{http://arxiv.org/abs/#1}{{\cob arXiv:#1}}}

\newcommand{\book}[5]{\emph{#1} (#2, #3, #5)}
\newcommand{\doin}[6]{\href{http://dx.doi.org/#1}{\cob #2\ #3 {\bf #4}, #5 (#6)}}

\newcommand{\doinn}[5]{\href{http://dx.doi.org/#1}{{\cob #2 {\bf #3}, #4 (#5)}}}
\newcommand{\doij}[5]{\href{http://dx.doi.org/#1}{{\cob #2 {\bf #3}, #4 (#5)}}}

\newcommand{\tia}[1]{{#1}.}

\begin{document}

\title{Fate of false vacuum in non-perturbative regimes: \it{Gravity effects}}

\author{Gianluca Calcagni\orcidC{}}
\email{g.calcagni@csic.es}
\affiliation{Instituto de Estructura de la Materia, CSIC, Serrano 121, 28006 Madrid, Spain}

\author{Marco Frasca\orcidA{}}
\email{marcofrasca@mclink.it}
\affiliation{Rome, Italy}

\author{Anish Ghoshal\orcidB{}}
\email{anish.ghoshal@fuw.edu.pl}

\affiliation{Institute of Theoretical Physics, Faculty of Physics, University of Warsaw, ul. Pasteura 5, 02-093 Warsaw, Poland}

\begin{abstract}
A recent analysis of the false-vacuum decay in non-perturbative regimes is here extended in the presence of Einstein gravity, computing the corresponding effective potential and decay rate. We consider a $\lambda \phi^4$ scalar field theory and we observe that, in comparison to the usual perturbative decay rate, the higher the coupling $\lambda$, the greater the decay probability. We evaluate the running of the self-interaction coupling and obtain a weakly coupled theory at lower energies, proving that Einstein gravity grants an even more reliable weak coupling approximation with the universe cooling down. We also provide an extended study of a non-minimal coupling $\xi$ of the scalar field with gravity showing how this term makes the false-vacuum decay more difficult. Minima can also disappear at large coupling $\xi$. We discuss possible applications of these results to cosmological phase transitions, gravitational-wave astronomy, and condensed matter systems.
\end{abstract}

\maketitle

\section{Introduction}

The recent blooming of gravitational-wave astronomy \cite{LIGOScientific:2016lio} and the
% Changed after referee's comments
% proof 
hints
of Higgs metastability after the detection of the Higgs boson at the LHC 
% Added last reference after referee's comments
\cite{Degrassi:2012ry,Isidori:2001bm,Hiller:2024zjp}
showed that the observation of the stochastic gravitational-wave background produced from primordial first-order phase transition could be a concrete possibility (see, e.g, \cite{LISACosmologyWorkingGroup:2022jok} and references therein). Some constraints on such a background were provided largely before the first observations of gravitational waves \cite{2009Natur.460..990A}, and they have been improved for different frequencies and parameters in the space of models in a systematic way \cite{LISACosmologyWorkingGroup:2022jok}. Such a swift evolution of this area of research has required the need of precise computations of the false-vacuum decay rate for all the available beyond-the-Standard-Model theories, with regimes ranging from weak to strong.

The technique to describe the decay of metastable states is the Euclidean path integral approach in the semi-classical approximation.

Callan and Coleman gave a proof of a decay of the vacuum through quantum fluctuations leading to spontaneous nucleation in a bubble of the true-vacuum phase \cite{Coleman:1977py,Callan:1977pt,Coleman:1980aw}. Such a nucleation rate is exponentially small, however, because the probability of such a decay is proportional to the exponential of the Euclidean action computed to a given $O(4)$-symmetric trajectory between the false vacuum and the tunneling point. Such a solution is known as the \emph{Coleman--De Luccia instanton} or \emph{bounce} when a gravitational interaction is present. Numerical computations have been largely performed for such a non-linear physics \cite{Claudson:1983et,Kusenko:1995jv,Kusenko:1996jn,Dasgupta:1996qu,Moreno:1998bq,John:1998ip,Masoumi:2016wot,Espinosa:2018hue,Espinosa:2018voj,Chigusa:2019wxb}, and analytic attempts as well \cite{Linde:1981zj,FerrazdeCamargo:1982sk,lee,Dutta:2011rc,Kanno:2012zf,Aravind:2014pva,Guada:2020ihz}, that apply very well in the so-called thin-wall approximation corresponding to a sufficiently small energy difference between the true and the false vacuum  \cite{Coleman:1980aw,Garfinkle:1989mv,Linde:1981zj,lee,Kanno:2012zf,Eckerle:2020opg}.  Bounce solutions for the Standard Model (SM) Higgs \cite{Espinosa:2018voj,Espinosa:2020qtq,Espinosa:2020cgk,Espinosa:2021tgx} or general scalar fields non-minimally coupled to gravity \cite{Isidori:2007vm,Salvio:2016mvj} or in higher-order gravity \cite{Salehian:2018yoq,Vicentini:2022pra,Vicentini:2021qoo,Vicentini:2020lhm} have been in the focus of recent interest.

Callan and Coleman's saddle-point approximation and the corresponding extension to include radiative corrections \cite{Weinberg:1992ds} have been generally applied to theories with weak coupling, a regime very challenging to realize if an actual bounce solution could possibly exists beyond the tree level with such techniques. Going beyond weak perturbation theory towards strongly-coupled regimes, false-vacuum decay methods which are accurate, versatile, robust, and may be applicable to phase transitions and the generation of gravitational waves have been actively sought. In Ref.\ \cite{Croon:2021vtc}, such an attempt was made via a quasi-stationary effective action relying upon functional renormalization group techniques \cite{Wetterich:1992yh}.

In an earlier paper \cite{Frasca:2022kfy}, some of us proposed an alternative and novel technique to compute the false-vacuum decay rate in strongly-coupled theories. Using exact solutions to the
% Changed after referee's comments
% Higgs theory 
quartic scalar field theory
as well as analytic expressions of the Green's functions expressed through Jacobi elliptical functions \cite{Frasca:2015yva}, we computed the effective action via the partition function and found the false-vacuum decay rate. Such a technique has been applied to quantum chromodynamics, confinement, and hadronic contributions to the muon magnetic moment \cite{Frasca:2021yuu,Frasca:2021mhi,Frasca:2022lwp,Chaichian:2018cyv}, to the Higgs sector of the Standard Model \cite{Frasca:2015wva}, and to studies of dark energy in cosmology \cite{Frasca:2022vvp}. The results obtained in this way hold even in strongly coupled regimes and go beyond the usual applications to high-energy physics and condensed matter systems.
In this paper, we expand our previous analysis and, on one hand, note an increase of the instability of the false vacuum at strong coupling. On the other hand, we generalize and explore the effect of gravitational interactions on the vacuum decay rate computed in the non-perturbative regimes, with the goal of understanding how gravity may impact the effective potential, the false vacuum, and its corresponding bouncing solution. We assume a minimal coupling for the scalar field coupled to gravity.

The non-minimal coupling $\xi$ between a scalar field and the Ricci scalar curvature plays a very important role in the context of vacuum decay. Often this coupling is recounted as an ingredient for the renormalizability of the scalar field in curved spacetime. However, it can still be negligible or even zero at certain energy scales. This coupling lays the cornerstone of the famous Higgs inflation model  \cite{Ade:2015lrj,Bezrukov:2007ep}; see Refs. \cite{Herranen:2014cua,Herranen:2015ima} for the impact of such non-minimal coupling on inflationary backgrounds, particularly if the scalar is the SM Higgs boson \cite{Isidori:2007vm,Rajantie:2016hkj}.
Since in this paper we work with theories with a single scalar field and a renormalizable potential, we will see that there are parameters controlling the influence of gravity on false-vacuum decay. We vary not only $\xi$ but also other parameters such as the self-quartic coupling. Tunneling both close and far from the thin-wall regime is discussed. We will discuss various scenarios involving quantum tunneling under the influence of gravity with and without non-minimal couplings. This can have applications such as preventing catastrophes in phenomenological theories, during inflation \cite{DiVita:2015bha,Bezrukov:2014ipa} or in the context with the matter-antimatter asymmetry problem addressed via baryogenesis \cite{Lewicki:2016efe}, or even to studies involving the string-theory landscape \cite{Demetrian:2005sr}.

%The paper is organized as follows. 
The paper has the following organization. In Sec.\ \ref{revi}, we recall an iterative procedure to obtain $n$-point correlation functions in quantum field theory (Sec.\ \ref{Dyson}) and, from these, the effective potential (Sec.\ \ref{effpo}). The non-perturbative method is presented in Sec.\ \ref{nonpe} in the absence of gravity. Apart from reviewing past results, we make a new comparison with the perturbative result and discuss the consequences of working at strong coupling. The modifications due to gravity are calculated in Sec.\ \ref{gravy} and, for
% Added after referee's comments
the case of non-minimal coupling, they are analyzed in Sec.\ \ref{gravy2}. Section \ref{disc} is devoted to discussion.

%\medskip

\section{Correlation functions and effective potential in scalar field theory}\label{revi}

When considering a Lorentzian spacetime, we work with the signature $(+,-,-,-)$, as in this section. Since we are interested in solutions interpolating nonequivalent vacua in the quantum theory, we will later move to the Euclidean signature $(+,+,+,+)$.

%\subsection{Dyson--Schwinger equations}
\subsection{Equations of Dyson--Schwinger}
\label{Dyson}

In this Section, we present an exact solution to the quantum field theory of the scalar field recently published  \cite{Frasca:2015yva,Frasca:2021yuu,Frasca:2021mhi,Frasca:2022lwp,Chaichian:2018cyv}. Such exact solution is not unique, but yields anyway a very precise account of the spectrum  \cite{Frasca:2017slg} and the beta function \cite{Chaichian:2018cyv} when properly extended to the Yang-Mills theory. The technique arises from a proposal due to Bender, Milton and Savage in another context \cite{Bender:1999ek} that permits to maintain the Dyson-Schwinger equations into their PDE shape. In this way, we can identify an exact solution for the 1P-correlation function that takes the form of a Fubini-Lipatov solution  \cite{Fubini:1976jm,Lipatov:1976ny} that represents a non-trivial vacuum with translation invariance broken.

We only report here the relevant formulas we will need in the following. For the interested reader, an extended discussion can be found in \cite{Frasca:2022kfy}. We consider a $\phi^4$ theory with action
\be
S=\int d^4x\left[\frac{1}{2}(\partial\phi)^2-\frac{\lambda}{4}\phi^4\right].
\ee
This yields the set of first Dyson--Schwinger equations
\be
\partial^2G_1(x)+\lambda[G_1(x)]^3+3\lambda G_2(x,x)G_1(x)+G_3(x,x,x)=0,
\ee
and
\begin{widetext}
\be
\partial^2G_2(x,y)+3\lambda[G_1(x)]^2G_2(x,y)+
3\lambda G_3(x,x,y)G_1(x)
+3\lambda G_2(x,x)G_2(x,y)
+G_4(x,x,x,y)=
\delta^4(x-y).
\ee
\end{widetext}

We see that higher-order correlation functions enter into the equations of the lower-order ones but evaluated at specific space-time points as $G_3(x,x,x)$, $G_4(x,x,x,y)$ and so on. If we select this special values to zero, we are really obtaining a Gaussian solution to the Dyson-Schwinger set of equations \cite{Frasca:2023uaw}.  These equations can be solved exactly yielding
\begin{equation}\label{solG1}
G_1(x)=\sqrt{\frac{2(p^2-m^2)}\lambda}\operatorname{sn}\left(p\cdot x+\theta,\kappa\right)
\end{equation}
with
\begin{equation}
p^2=\frac12\left(\sqrt{m^4+2\lambda\mu^4}+m^2\right),\qquad
\kappa=\frac{m^2-p^2}{p^2}
\end{equation}
where $\mu$ and $\theta$ are constants arising from the integration
%integration constants 
and $\operatorname{sn}(\zeta,\kappa)$ is Jacobian elliptic function. 
%of the first kind. 
We just defined
\be
m^2=3\lambda G_2(0)
\ee
for the gap equation due to quantum effects. Such a term needs renormalization being
\be
m^2=3\lambda\int\frac{d^4p}{(2\pi)^4}G_2(p).
\ee
In momentum space, the two-point function is given by
\be
\label{eq:G2e}
G_2(p)=\frac{\pi^3}{2\sqrt{\kappa}(1-\kappa)K^3(\kappa)}
\sum_{n=0}^\infty(-1)^n(2n+1)^2\frac{q^{n+\frac{1}{2}}}{1-q^{2n+1}}\frac{1}{p^2-m_n^2+i\epsilon}\,,
\ee
where $K(\kappa)$ is the complete elliptic integral of the first kind such that $K^*(\kappa)=K(1-\kappa)$, 
\be
\kappa=\frac{m^2-\sqrt{m^4+2\lambda\mu^4}}{m^2+\sqrt{m^4+2\lambda\mu^4}},
\ee
and
\be
q=e^{-\pi\frac{K^*(\kappa)}{K(\kappa)}}\,.
\ee
The spectrum is given by
\be
m_n=(2n+1)\frac{\pi}{2K(\kappa)}\frac{1}{\sqrt{2}}\sqrt{m^2+\sqrt{m^4+2\lambda\mu^4}}.
\ee
Therefore, assuming the shift $m^2$ small enough, the equations simplify to
\be
\label{eq:G2s}
G_2(p)=\frac{\pi^3}{4K^3(-1)}
\sum_{n=0}^\infty(2n+1)^2\frac{e^{\left(n+\frac{1}{2}\right)\pi}}{1+e^{-(2n+1)\pi}}\frac{1}{p^2-m_n^2+i\epsilon}\,,
\ee
where now
\be
m_n=(2n+1)\frac{\pi}{2K(-1)}\left(\frac{\lambda}{2}\right)^\frac{1}{4}\mu.
\ee

Similarly, the three-point and four-point functions are given by
\begin{equation}
\label{eq:G_3}
   G_3(x,y,z)=-6\lambda\int dx_1 G_2(x-x_1)G_1(x_1-y)G_2(x_1-y)G_2(x_1-z),
\end{equation}
and
\begin{eqnarray}
    G_4(x,y,z,w)&=&-6\lambda\int dx_1 G_2(x-x_1)G_2(x_1-y)G_2(x_1-z)G_2(x_1-w) \nonumber\\ \nonumber
    &&-6\lambda\int dx_1G_2(x-x_1)\left[G_1(x_1-y)G_2(x_1-y)G_3(x_1-z,x_1-w)\right. \\ 
    &&\left.+G_1(x_1)G_2(x_1-z)G_3(x_1-y,x_1-w)
    +G_1(x_1-y)G_2(x_1-w)G_3(x_1-y,x_1-z)\right].
\end{eqnarray}
We can see that 
\be
G_3(x,x,x)=-6\lambda\int dx_1 G_2(x-x_1)G_1(x_1-x)G_2(x_1-x)G_2(x_1-x)
\ee
where the product $G_2(x-x_1)G_2(x_1-x)$ enters into the integral. This grants that such particular value of the $G_3$ function is zero as promised.

%% Added after referee's comments on 02-09-2022
A few words are needed about the breaking of translation invariance in the correlation function $G_1$ in eq.\ (\ref{solG1}). It must be emphasized that this is not an observable of the theory and such a breaking cannot be seen anywhere by any kind of experiment. This is due to the two-point function recovering translation invariance so that the Lehman-Symanzik-Zimmermann (LSZ) theorem shows that, whatever scattering or decay amplitude one can computes, no observation is possible of breaking of translation invariance.

There is no way to perform a physical measurement to obtain $G_1$ by experiments and so, practically, no breaking of translation invariance can ever be seen. Last but not least, such theories display some form of confinement, limiting even more the range of such a violation.
%%

% Changed after referee's comments on 05/03/2023
\subsection{
%Effective potential
Effective action
}\label{effpo}

We evaluate the effective potential, defined in  \cite{Coleman:1973jx}, using the exact solution we discussed above. Our main reference is \cite{Frasca:2022kfy} and the partition function can be written down as
\be
Z[j]=\sum_{n=1}^\infty\left(\prod_{m=1}^n\int d^4x_m\right) G_n(x_1..x_n)
\prod_{p=1}^nj(x_p),
\ee
%\ag{Please define G.}
%with 
where
$G_n$ 
%being 
is the $n$-point function,
%From this, one is able to get, in principle, all the $n$-point functions exactly. 
Given this equation, all the nP-correlation functions can be computed exactly, in principle.
% Added after referee's comments on 08/03/2023
In our case, 
%in principle, 
we could have a closed analytical formula for all the correlation functions and so, the partition function is known exactly aside from computational complexity. Then, we move to derive the relative expressions for n-point vertexes.
The effective action can be obtained starting with 
\be
\phi_c(x)=\langle\phi(x)\rangle=\frac{\delta W[j]}{\delta j(x)}\,,
\ee
where
\be
W[j]=\ln Z[j].
\ee
Therefore, by a Legendre transform, we get the effective action $\Gamma$ as
%is given by a Legendre transform
\be
\Gamma[\phi_c]=W[j]-\int d^4xj(x)\phi_c(x),
\ee
giving
\be
\Gamma[\phi_c]=\sum_{n=1}^\infty\left(\prod_{m=1}^n\int d^4x_m\right) \Gamma_n(x_1..x_n)
\prod_{p=1}^n\phi_c(x_p).
\ee

The $n$-point vertices and correlation functions that interest us are given by
\begin{widetext}
\bea
\label{eq:Gn}
G_2(x_1,x_2)&=&\Gamma_2^{-1}(x_1,x_2), \nonumber \\
G_3(x_1,x_2,x_3)&=&\int d^4x'_1d^4x'_2d^4x'_3\Gamma_3(x'_1,x'_2,x'_3)G_2(x_1,x'_1)G_2(x_2,x'_2)G_2(x'_3,x_3).
\eea
\end{widetext}
It also holds
\begin{widetext}
\be
\int dz\frac{\delta^2\Gamma[\phi_c]}{\delta\phi_c(x)\delta\phi_c(z)}
\left(\frac{\delta^2\Gamma[\phi_c]}{\delta\phi_c(z)\delta\phi_c(y)}\right)^{-1}
=\delta^4(x-y).
\ee
\end{widetext}
Finally, the formula we will use is
\begin{widetext}
\be
\label{eq:EA}
%L^4V_{\rm eff}[\phi_c]
%\int d^4xV_{\rm eff}[\phi_c]
\Gamma[\phi_c]=-\sum_n\frac{1}{n!}\int d^4x_1\ldots d^4x_n\Gamma_n(x_1,\ldots,x_n)
(\phi_c(x_1)-\phi_0(x_1))\ldots(\phi_c(x_n)-\phi_0(x_n)),
\ee
\end{widetext}
%where $L^{4}$ is 
where the integration is intended on all
the volume of spacetime, which we will assume finite.

In our case is
\be
\left.\phi_c(x)\right|_{j=0}=\phi_0(x).
\ee
 
This yields the 1P-correlation function representing a non-trivial vacuum, having retained the space-time dependence. Indeed, standard perturbation theory starts with $\phi_0=0$.

\medskip

%\section{Effective potential}

\section{Non-perturbative method for false-vacuum decay}\label{nonpe}

In an optic of a self-contained paper, we show here the computations given in Ref.\cite{Frasca:2022kfy}. This will yield a bounce action obtained by a stationary phase approximation with the solutions of the equation of motion given in an Euclidean space. For our case, one has
\begin{equation}
    S[\phi] = -\int d^4 x \left\{\frac{1}{2} [\partial \phi(x)]^2 + V[\phi(x)]\right\}.
\end{equation} 

Then, the decay rate $\gamma$ for the false vacuum, per unit volume, is given by
\begin{align}
%\be
    \gamma = - 2\, {\rm Im} \,\mathcal{E}, \qquad e^{- L^4\mathcal{E}} \simeq Z = \int_{\phi_{\rm FV}}^{\phi_{\rm FV}} D \phi\, e^{- S[\phi]},
\end{align}

where $\phi_{\rm FV}$ is the false vacuum state and the path integral is evaluated assuming $\phi(x,\pm\infty) = \phi_{\rm FV}$. The integral is given by a sum over the saddle (stationary) points that are solutions to the equation of motion  \cite{Coleman:1977py,Callan:1977pt}. In our case, by \emph{bounce solution} we simply mean a solution such that the two-point correlation function has the property to decay at infinity being an infinite sum of Yukawa propagators. Indeed, from (\ref{eq:G2e}) one has
\be
G_2(p)=\sum_{n=0}^\infty\frac{B_n}{p^2-m_n^2+i\epsilon}\,,
\ee
that can be transformed back to position space:
\be\label{Gxy}
G_2(x,y) = \begin{cases}
-\frac{1}{4 \pi} \delta(s) + \sum_{n=0}^\infty\frac{m_n}{8 \pi \sqrt{s}} H_1^{(1)}(m_n \sqrt{s}) & s \geq 0 \\ -\sum_{n=0}^\infty\frac{i m_n}{ 4 \pi^2 \sqrt{-s}} K_1(m_n \sqrt{-s}) & s < 0,\end{cases} 
\ee
where $s= (x^0 - y^0)^2 - (\bm{x} - \bm{y})^2$ and $H_1^{(1)}(x)$ and $K_1(x)$ are the Hankel and the Bessel modified function, respectively. 
%Use has been made of the identity $\sum_{n=0}^\infty B_n=1$. 
We used the identity $\sum_{n=0}^\infty B_n=1$.
From this equation, the proper asymptotic behavior can be obtained and one can see that (\ref{Gxy}) is suppressed at $s\to\pm\infty$.

We have not considered radiative corrections, maintaining a tree level action. On the other hand, whenever radiative corrections become relevant for the tree-level action, we are considering \textit{bounce-like} field configuration that are not saddle points of it. Then, the phase will dominate the integral anyway as its variation will be slow enough \cite{Iliopoulos:1974ur}.

As stated elsewhere, we are considering an exact solution that represents a Gaussian solution to the set of Dyson-Schwinger equations. This has some important implications about radiative corrections as seen in calculation of the effective potential using perturbation theory and also in the missing of corrections to the kinetic term of the action. We also note that our exact non-perturbative effective action is not related to the bounce formalism.

Given the effective potential, we can compute the decay rate of the false vacuum in agreement to Refs. \cite{Coleman:1977py,Callan:1977pt}. 
The decay rate is 
%given by
% Modified after referee's comments
\be
%\gamma\sim L^{-4}e^{-\Gamma[\phi_c]},
\gamma\sim e^{-B},
\ee

where the $B$ exponent is computed below through the effective action, in agreement with Ref.\cite{Coleman:1977py,Callan:1977pt}.

Using the solutions for the 1P- and 2P-correlation functions, one gets
\begin{widetext}
\bea
G_3(x-x_1,x-x_2)&=&-\int d^4x_3G_2(x-x_3)V'''[\phi_0(x_3)]G_2(x_1-x_3)G_2(x_2-x_3) \nonumber \\
&=&-\int d^4x_1d^4x_2d^4x_3G_2(x-x_3)V'''[\phi_0(x_3)]\delta^4(x-x_1)\delta^4(x-x_2)G_2(x-x_3)G_2(x-x_3),
\eea

from which, by comparison with Eq.\ (\ref{eq:Gn}), we get
\be
\Gamma_3(x_1,x_2,x_3)=-V'''[\phi_0(x_1)]\delta^4(x_1-x_2)\delta^4(x_1-x_3).
\ee
\end{widetext}

Then, we have to evaluate for the effective action
\be
\label{eq:FullU}
U(\phi_c)=V(\phi_c)+\frac{1}{2}V_2(\phi_c-\phi_0)^2-\frac{1}{3!}|V_3|(\phi_c-\phi_0)^3.
\ee

From eq.(\ref{eq:G2s}) we get $G_2(p)$ and the mass spectrum.
 
This yields immediately for the 2P-vertex function
\be
V_2=\frac{K(i)}{2\pi}\mathscr{A}\sqrt{2\lambda}\mu^2=\sqrt{2\lambda}\mu^2,
\ee
with\footnote{The following identity also holds: $\left[\sum_{n=0}^\infty (2n+1)^2A_n\right]^{-1}=4K^3(i)/\pi^3$.}
\be
%\mathscr{A}=\left(\sum_{n=0}^\infty A_n\right)^{-1} = 4.792560938942369\ldots=\frac{2\pi}{K(i)},
\mathscr{A}=\left(\sum_{n=0}^\infty A_n\right)^{-1}=\frac{2\pi}{K(i)},\qquad
A_n=\frac{e^{-\left(n+\frac{1}{2}\right)\pi}}{1+e^{-(2n+1)\pi}}.
\ee
Then, one has
\be
V_3=-6\lambda\phi_0(0)=-3\mu(2\lambda)^\frac{3}{4}\operatorname{sn}(\theta,i),
\ee
%Here, $\theta$ is an integration constant of the equation for the one-point correlation function. 
where $\theta$ arises as an integration constant from the equation of the 1P-correlation function. This shows that the $\mathbb{Z}_2$ symmetry, that holds at a classical level, is indeed broken by quantum corrections.
%% Modified after referee's comments on 02-09-2022
There is a deep reason for this in our formalism as such an effect 
%arises 
is
%due to 
a consequence of
the presence of a zero mode in the theory. We discuss it in Appendix \ref{appA}.
%%

%Now we can write the Euclidean effective action as
The Euclidean effective action 
%can be written down as
is then given by
\be
\Gamma_E(\phi_c)=-\int d^4x\left[\frac{1}{2}(\partial\phi_c)^2+U(\phi_c)\right]
\ee
%with 
with the effective potential
\be
\label{eq:Ueff}
U(\phi_c)=
\frac{1}{2}V_2(\phi_c-\phi_0)^2+\frac{1}{3!}V_3(\phi_c-\phi_0)^3+\frac{\lambda}{4}\phi_c^4.
\ee

Here, we can neglect the contribution due to  $\phi_0$ by taking the formal limit $\lambda\rightarrow\infty$ 
%as it goes like 
given the dependence on
$\lambda^{-\frac{1}{4}}$. Thus, the extremes of the potential in the strong coupling limit are given by
\be
2^\frac{3}{4}k_1\mu\lambda^{-\frac{1}{4}}, \quad 0, \quad -2^\frac{3}{4}k_2\mu\lambda^{-\frac{1}{4}},
\ee
where $k_1=k_1(\theta)$ and $k_2=k_2(\theta)$ are
%numerical constants depending on the initial phase $\theta$ given by
just numbers parameterized by the initial phase $\theta$ and are given by
\bea
k_1(\theta)&=&3\operatorname{sn}(\theta,i) + \sqrt{9\operatorname{sn}^2(\theta,i) - 8}, \nonumber \\
k_2(\theta)&=&
-3\operatorname{sn}(\theta,i) + \sqrt{9\operatorname{sn}^2(\theta,i) - 8}.
\eea

Such extremes exist provided
\be
\label{eq:lim}
|\operatorname{sn}(\theta,i)|>\sqrt{\frac{8}{9}}\approx 0.94.
\ee

Thus, it is enough to take $|\operatorname{sn}(\theta,i)|$ near 1.

\begin{figure}[H]
\centering
\includegraphics[height=8cm,width=10cm]{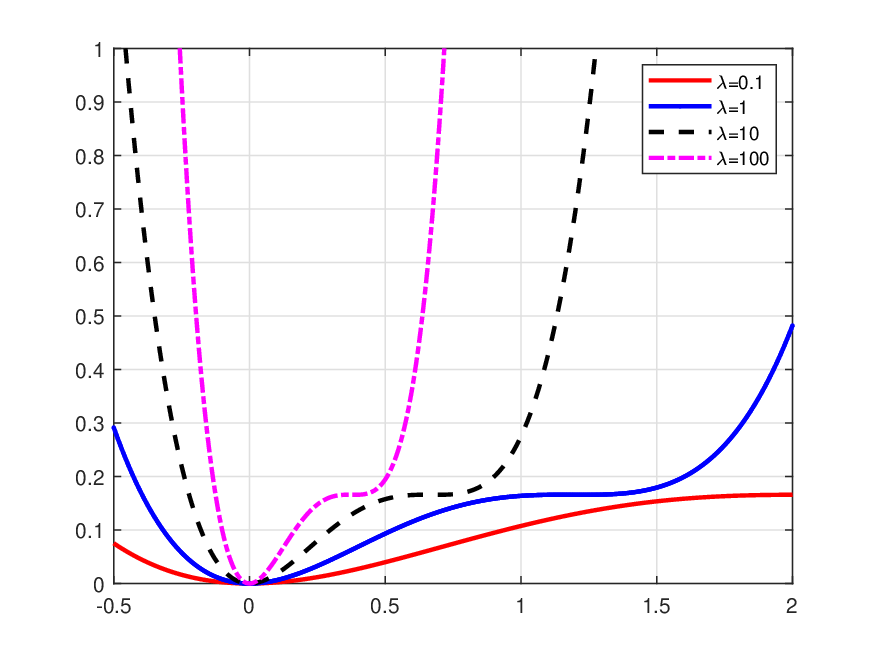}
\caption{\it 
%Plot of $U(\phi)$ as given in Eq.~(\ref{eq:FullU}) for a fixed constant $\phi_0$.
Plot of $U(\phi)$ obtained from Eq.~(\ref{eq:FullU}) at constant $\phi_0$.
The effect of non-perturbative regimes is noted by the way the decay is eased at increasing $\lambda$.
%\ag{Let us think what is the physical meaning of spreading of curve ?}
\label{fig2} 
}
\end{figure}
\begin{figure}[H]
\centering
\includegraphics[height=8cm,width=10cm]{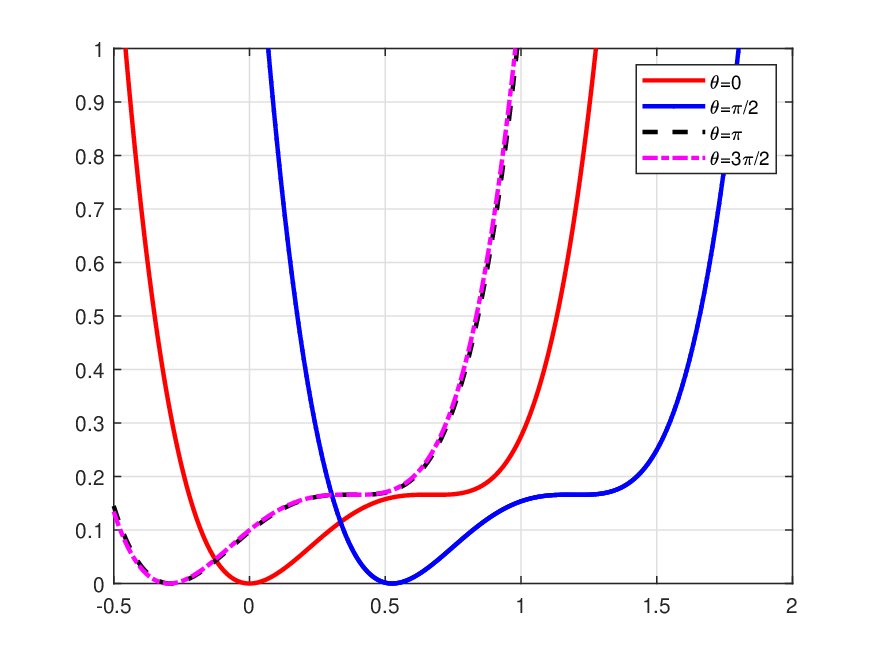}
\caption{\it 
%Plot of $U(\phi)$ at fixed $\lambda$ but varying the phase $\theta$.
\label{fig3} 
Same as in Fig.~\ref{fig3} but with varying $\theta$ and keeping $\lambda$ fixed.
}
\end{figure}

 
Similarly, setting $\lambda=10$ and varying the phase we get the plot in Fig.~\ref{fig3}. Here the effect is a simple horizontal shift of the potential without altering its shape. %\mf{Correct. Inserted in the main text.}

To evaluate the decay rate, we use the argument presented in Refs.\ \cite{Coleman:1977py,Weinberg:1996kr}

\be
B=\frac{27\pi^2\mathscr{S}^4}{2\epsilon^3}
\ee
with
\be\label{Seps}
\mathscr{S}=\int_0^{\langle\phi_c\rangle}df\sqrt{2U(f)}\,,\qquad
\epsilon = U(\langle\phi_c\rangle).
\ee
We have $\epsilon = c_1\mu^4$ with $c_1$ being a numerical constant. For $\lambda$ large enough, we can evaluate $\mathscr{S}$ as seen in Ref.\cite{Frasca:2022kfy} obtaining

\be
\mathscr{S}\simeq\frac{1}{2}\sqrt{2U(\langle\phi_c\rangle)}\langle\phi_c\rangle=\frac{1}{2}\sqrt{2\epsilon}\langle\phi_c\rangle.
\ee
%Assuming 
Taking $\rm{sn}(\theta,i)\approx 1$, we get
\be\label{epsil}
%\epsilon\approx\mu^4/8,
\epsilon=\kappa_2(\theta)\frac{\mu^4}{16},
\ee
where
\begin{widetext}
\be
\kappa_2(\theta)=
-27\operatorname{sn}^4(\theta,i) + 9\sqrt{9\operatorname{sn}^2(\theta,i) - 8}\,\operatorname{sn}^3(\theta,i) + 36\operatorname{sn}^2(\theta,i) - 8\sqrt{9\operatorname{sn}^2(\theta,i) - 8}\,\operatorname{sn}(\theta,i) - 8.
\ee
\end{widetext}
Then,
\be
\label{eq:B1}
%B=\frac{27\pi^2}{8\lambda}.
B=\frac{432\pi^2}{\lambda}\frac{k_2^4(\theta)}{\kappa_2(\theta)}\,.
\ee

For $\operatorname{sn}(\theta,i)=1$, one has
%we get the formula
\be
\label{eq:B2}
B=3456\frac{\pi^2}{\lambda},
\ee
holding for $\lambda\gg 1$. Then, we obtain finally for the decay rate
\be \label{eq:B21}
\gamma\sim L^{-4}e^{-\frac{432\pi^2}{\lambda}\frac{k_2^4(\theta)}{\kappa_2(\theta)}}
\simeq L^{-4}e^{-3456\frac{\pi^2}{\lambda}}.
\ee

We have assumed $\operatorname{sn}(\theta,i)=1$ satisfying the given condition $|\operatorname{sn}(\theta,i)|>\sqrt{8/9}\approx 0.94$.

This result should be compared with the one given in standard textbooks in the weak perturbation limit \cite{Weinberg:1996kr}, in order to understand where the crossing border lies,
\be \label{eq:comp1}
B_{\rm weak}=\frac{8\pi^2}{3\lambda}.
\ee
We see that
\be \label{eq:comp2}
B= 1296\,B_{\rm weak}\,.
\ee

This defines a crossing line between the strong- and weak-coupling regimes.

\section{Corrections due to gravity}\label{gravy}

The effect of gravity in the thin-wall approximation can be easily evaluated by considering Ref.~\cite{Coleman:1980aw}. Assuming Einstein gravity with standard Einstein equations $R_{\mu\nu}-(1/2) g_{\mu\nu}R=8\pi G T_{\mu\nu}$, the physical interpretation and the arguments leading to the non-perturbative result are the same and the only difference is in the shape of the potential. One starts with a universe sitting in the false vacuum of the scalar field. Through quantum tunneling, a bubble of true vacuum of radius $\rho_0$ is created in space within the false vacuum. The bubble expands to incorporate all nearby regions, eventually converting the whole universe to the true vacuum. %\mf{Inserted into the main text.}

\subsection{Decay rate and bubble radius}

In the absence of gravity, a calculation identical to that in Ref.~\cite{Coleman:1980aw} gives the bubble radius at the moment of its materialization
% Moved \Lambda to \kappa_0 to avoid confusion with the cosmological constant
\be\label{rho0}
{\bar\rho}_0=\frac{3\mathscr{S}}{\epsilon}\,,
\ee
with $\mathscr{S}$ and $\epsilon$ given by Eq.\ (\ref{Seps}). Switching gravity on, when the decay is from a minimum at positive energy to 0 the radius becomes smaller,
\be
{\bar\rho}=\frac{\bar\rho_0}{1+\frac{{\bar\rho}_0^2}{4\kappa_0^2}}.
\ee
while
\be
\label{comp:gr1}
B_g=\frac{B}{\left[1+\frac{{\bar\rho}_0^2}{4\kappa_0^2}\right]^2}<B\,,
\ee
where
\be \label{Lam}
\kappa_0 = \left(\frac{8\pi G\epsilon}{3}\right)^{-\frac{1}{2}},
\ee
being $G$ Newton's constant. In this case, gravity reduces $B_g<B$ and makes the decay (creation of the true-vacuum bubble) more likely. In contrast, when the decay is from 0 to a negative-energy minimum, the initial radius of the bubble is shrunk,
\be 
{\bar\rho}=\frac{\bar\rho_0}{1-\frac{{\bar\rho}_0^2}{4\kappa_0^2}},
\ee
and the decay becomes less likely:
\be \label{comp:gr2}
B_g=\frac{B}{\left[1-\frac{{\bar\rho}_0^2}{4\kappa_0^2}\right]^2}>B\,.
\ee

Using the results of Sec.\ \ref{nonpe}, we see that the impact of the non-perturbative regime is to further enhance the decay rate from a positive-energy false vacuum to 0 and to compensate the slower decay from a zero-energy false vacuum to the negative-energy true vacuum. One can check this explicitly incorporating the results of \cite{Isidori:2007vm}, computing $\bar\rho$ using the potential (\ref{eq:Ueff}).

At this point, we can also find a condition on $\lambda$ such that $B_g^{\rm strong}=B_{\rm weak}$. In fact, the left-hand side includes both gravitational and non-perturbative effects, while the right-hand side includes none. In this way, we determine for which value of $\lambda$ non-perturbative and gravity effects compensate each other. Plugging Eq.\ (\ref{eq:comp2}) into (\ref{comp:gr2}) and setting $B_g=B_{\rm weak}$ as argued above, we obtain
\be 
\frac{1296}{\left[1-\frac{{\bar\rho}_0^2}{4\kappa_0^2}\right]^2}=1\,,
\ee
which yield the running coupling
\be\label{lava}
\sqrt{\lambda}=\frac{9}{35}2^\frac{9}{2}(\kappa_0\mu)^{-2}.
\ee
In order to obtain (\ref{lava}), we combined Eqs.\ (\ref{epsil}), (\ref{eq:comp2}), (\ref{rho0}) and (\ref{Lam}). We just point out that $(\Lambda\mu)^{-2}\propto \mu^2/M_{\rm Pl}^2$, where $M_{\rm Pl}$ is the Planck mass. We can recognize here the running coupling law for the scalar field already put forward in \cite{Frasca:2013tma}. 
%This law characterizes also conformally flat general relativity \cite{Frasca2022mod}. 
We get a infrared weakly coupled theory in the limit $\mu\rightarrow 0$. 
% Removed not so well-motivated statement.
%Indeed, from the coupling of the $\phi^4$ term in the one-loop effective action, the beta function is immediately given by
%\be 
%\beta(\lambda)=4\lambda.
%\ee
%
The conclusion is that, if the theory is weakly coupled in the infrared, then perturbative quantum field theory is an adequate tool to describe the physics at large scales. In this case, if the coupling takes the value (\ref{lava}), then gravity effects exactly compensate  non-perturbative corrections.

As a side note not relevant for the rest of the analysis, we can interpret the scalar field as arising from a conformal transformation of the 
% Changed after referee's comments
% Einstein--Hilbert 
Starobisnky
action \cite{Momeni:2025ylx}, in vacuum. 

Moving to the Einstein frame, we get a minimally coupled scalar field with the Einstein-Hilbert action.

\subsection{Instabilities}

In \cite[Section III]{Lee:2014uza}, the presence of instabilities was noted with a general argument independent of the shape of $U''$. Therefore, the same argument applies here. An explicit check can be carried on noting that the eigenvalue equation
\begin{equation}\label{eigeneq}
[-\Box + U''(\phi_c)]\, \eta = A\, \eta
\end{equation} 
implies the eigenvalue equation $-\Box \eta = (A-U'') \eta$, where $-\Box + U''$ is the Hessian in Euclidean signature, $\eta$ is the eigenfunction and $A$ is the eigenvalue. Thanks to the spherical symmetry of the problem, we can write $\Box=d^2/ds^2+(3/s)d/ds$ and consider single-variable eigenfunctions $\eta(s)$. Equation (\ref{eigeneq}) is real-valued provided $A$ and $U''$ are real, which is true for ${\rm Im}[{\rm sn}(\theta,i)]=0$. This selects a discrete set of values for the parameter $\theta$. Taking such values and assuming $A < 0$, one can check that, since $-U''$ has concavity pointing downwards, it is impossible to get $A-U''>0$ for all values of $\phi_c$. To state it differently, for $A < 0$ the eigenfunction $\eta(s)$ is well defined.

\section{The case of the non-minimal coupling}
\label{gravy2}

From Ref.~\cite{Czerwinska:2016fky}, we assume the following model for a de Sitter geometry:
\begin{equation}
\label{eq:lagrangian}
\mathcal{L}= 
\frac{1}{2}(\partial \phi )^2-V(\phi)+\frac{1}{2}\frac{\mathcal{R}}{\kappa}\left(1-\xi \frac{12}{\Lambda} \phi^2 \right),
\end{equation}
with
\begin{equation}
\label{eq:potential}
V(\phi)= -\frac{1}{4} a^2 (3 b - 1) \phi^2 + \frac{1}{2} a (b - 1) \phi^3 + \frac{1}{4} \lambda\phi^4+a^4 c,
\end{equation}
%\ag{Need to define $\kappa$, $\mathcal{R}$, $\Lambda$, $\xi$.}
where $\phi$ is the scalar field, $\kappa=8\pi G$ being $G$ the Newton constant, $\mathcal{R}$ is the Ricci scalar, $\Lambda$ is the cosmological constant and $\xi$ is the non-minimal coupling. In our case, referring to \cite{Frasca:2022kfy}, 
%we will have \ag{physica reason explanation what are these}
we can always cast the given potential in a standard form with the choices
\be
a=\frac{1}{4}(-V_3+\sqrt{V_3^2-4V_2}), \qquad b=\frac{3V_2-V_3^2-V_3\sqrt{V_3^2-4V_2}}{3V_2}, \qquad c=0.
\ee
% being
% \be
% V_2=\sqrt{2\lambda}\mu^2, \qquad V_3=3(2\lambda)^\frac{3}{4}\mu\operatorname{sn}(\theta,i).
% \ee
% Here $\mu$ and $\theta$ are arbitrary integration constants and sn is a Jacobi elliptical function. 
In this way, we are back to the potential given in \cite{Frasca:2022kfy} that arises from a non-perturbative analysis on a $\phi^4$ theory and is recast as
\be
V(\phi)=\frac{1}{2}V_2(\phi-\phi_0)^2-\frac{1}{3!}V_3(\phi-\phi_0)^3+\frac{1}{4}\lambda\phi^4.
\ee
In this expression, we added a background solution $\phi_0(x)$. One has $\phi_0\propto \lambda^{-\frac{1}{4}}$, so that this becomes increasingly small at large coupling as pointed out in \cite{Frasca:2022kfy}. We are considering metrics with $\mathcal{R} = {\rm const}$,  i.e., de Sitter, anti-de Sitter and Minkowski. %The latter is very well-known. The question is how the exact solutions for the scalar field change in this case.

%\medskip

\subsection{Dyson-Schwinger equations in non-minimal gravity}
\label{sec3}

{Starting from the Lagrangian (\ref{eq:lagrangian}), we can write down the set of Dyson-Schwinger equations \cite{Frasca:2015yva}
\bea
&&\partial^2 G_1(x)+\lambda\left([G_1(x)]^3+3G_2(x,x)G_1(x)+G_3(x,x,x)\right)+\xi R G_1(x)=0 \nonumber \\
&& \nonumber \\
&&\partial^2G_2(x,y)+\lambda\left(3[G_1(x)]^2G_2(x,y)+3G_2(x,x)G_2(x,y)+3G_3(x,x,y)G_1(x)+G_4(x,x,x,y)\right)+\xi R G_2(x,y)=\delta^4(x-y) \nonumber \\
&&\nonumber \\
&&\partial^2G_3(x,y,z)+\lambda\left[6G_1(x)G_2(x,y)G_2(x,z)+3G_1^2(x)G_3(x,y,z)+3G_2(x,z)G_3(x,x,y)+3G_2(x,y)G_3(x,x,z)\right. \\ \nonumber
&&\left.+3G_2(x,x)G_3(x,y,z)+3G_1(x)G_4(x,x,y,z)+G_5(x,x,x,y,z)\right]+\xi R G_3(x,y,z)=0 \\ \nonumber
&& \nonumber \\
&&\partial^2G_4(x,y,z,w)+\lambda\left[6G_2(x,y)G_2(x,z)G_2(x,w)+6G_1(x)G_2(x,y)G_3(x,z,w)+6G_1(x)G_2(x,z)G_3(x,y,w)\right.\\ \nonumber
&&+6G_1(x)G_2(x,w)G_3(x,y,z)+3G_1^2(x)G_4(x,y,z,w)+3G_2(x,y)G_4(x,x,z,w)+3G_2(x,z)G_4(x,x,y,w)  \\ \nonumber
&&\left.+3G_2(x,w)G_4(x,x,y,z)+3G_2(x,x)G_4(x,y,z,w)+3G_1(x)G_5(x,x,y,z,w)+G_6(x,x,x,y,z,w)\right]+\xi R G_4(x,y,z,w)=0 \\ \nonumber
&\vdots&,
\eea
where $\partial^2=(1/\sqrt{-g})\partial_\alpha\sqrt{-g}g^{\alpha\beta}\partial_\beta$ is the Laplace-Beltrami operator and $R=\frac{12\mathcal{R}}{\kappa\Lambda}$. We want to evaluate the 1P- and 2P-correlation functions for some choice of the metric. We consider the de Sitter case referring to \cite{Czerwinska:2016fky}. Our aim is to understand the behavior of the theory when the coupling $\lambda$ becomes large and we enter a non-perturbative regime. We choose flat slicing coordinates 
%We choose the static coordinates and we assume the presence of a cosmological horizon. 
and we can write the line element as
\be
ds^2=dt^2-e^{2\sqrt{\Lambda}t}[dr^2-r^2(d\theta^2+\sin^2\theta d\phi^2)],
\ee
where $\Lambda$ is the cosmological constant. This means that the Laplace-Beltrami operator takes the form
\be
\Box_{dS}=\partial_t^2+3\sqrt{\Lambda}\partial_t-e^{-2\sqrt{\Lambda}t}\Delta_2
%r^{-2}\partial_rr^2\left(1-\frac{\Lambda}{3}r^2\right)\partial_r-\frac{L^2}{r^2}
\ee
where $\Delta_2$ is the standard Laplacian in spherical coordinates. Therefore, for the 1P-correlation function one can write
\be
\label{eq:G11}
\Box_{dS} G_1(x) +\lambda\left([G_1(x)]^3+3G_2(x,x)G_1(x)+G_3(x,x,x)\right)+ 12\xi R G_1(x)=0.
\ee
To solve this equation we assume, as usual, that wherever an evaluation at equal position is performed in an $n$P-correlation function with $n>2$, such a contribution is zero. Therefore, $G_3(x,x,x)=0$ that is checked {\sl a posteriori}. Then, we introduce the mass term arising from quantum fluctuations as $m_q^2=3\lambda G_2(x,x)$ being $G_2(x,x)$ a finite constant after proper renormalization providing a dependence on both $\Lambda$ and $\xi$. We get
\be
\label{eq:G1_r}
\Box_{dS} G_1(x) +\lambda [G_1(x)]^3+M_q^2G_1(x)=0,
\ee
where
\be
\label{eq:Mq_0}
M_q^2=m_q^2+\xi R,
\ee
and, for the 2P-function,
\be
\label{eq:G2_r}
\Box_{dS} G_2(x,x') +3\lambda [G_1(x)]^2G_2(x,x')+M_q^2G_2(x,x')=-ie^{-3\sqrt{\Lambda}t}\delta^4(x-x').
\ee
Finally, we assume the ordering $\lambda\gg 1$ and rescale the time variable as $t\rightarrow\sqrt{\lambda}t$ and expand
\be
G_1=\sum_{n=0}^\infty\lambda^{-n/2}G_1^{(n)}.
\ee
We observe that also the mass term in Eq.\ (\ref{eq:G1_r}) is $O(\lambda)$. We get the following gradient expansion
\bea
\label{eq:series}
&&\partial_t^2 G_1^{(0)}(x)+\lambda [G_1^{(0)}(x)]^3
+M_q^2G_1^{(0)}(x)=0 \nonumber \\
&&\partial_t^2 G_1^{(1)}(x)+3\lambda [G_1^{(0)}(x)]^2G_1^{(1)}(x)
+M_q^2G_1^{(1)}(x)=-3\sqrt{\Lambda}\partial_tG_1^{(0)}(x)\nonumber \\
&&\partial_t^2 G_1^{(2)}(x)+3\lambda [G_1^{(0)}(x)]^2 G_1^{(2)}(x)
+M_q^2G_1^{(2)}(x)
= e^{-2\sqrt{\Lambda}t}\Delta_2 G_1^{(0)}(x)-3\sqrt{\Lambda}\partial_tG_1^{(1)}(x) \nonumber \\
&&\partial_t^2 G_1^{(3)}(x)+3\lambda [G_1^{(0)}(x)]^2 G_1^{(3)}(x)
+M_q^2G_1^{(3)}(x)
= \nonumber \\
&&-\lambda[G_1^{(1)}(x)]^3-6\lambda G_1^{(0)}(x)G_1^{(1)}(x)G_1^{(2)}(x)
+e^{-2\sqrt{\Lambda}t}\Delta_2 G_1^{(1)}(x)-3\sqrt{\Lambda}\partial_tG_1^{(2)}(x) \nonumber \\
&&\vdots.
\eea
The leading order 1P-correlation function can be written as
\be
\label{eq:G10}
G_1^{(0)}(t,0)=\sqrt{\frac{2\mu^4}{M_q^2+\sqrt{M_q^4+2\lambda\mu^4}}}\operatorname{sn}\left(Et+\theta,\kappa\right),
\ee
where sn is a Jacobi elliptical function, 
$$\kappa=\sqrt{\frac{M_q^2-\sqrt{M_q^4+2\lambda\mu^4}}{M_q^2+\sqrt{M_q^4+2\lambda\mu^4}}},$$
and $E^2=M_q^2+\lambda\mu^4/(M_q^2+\sqrt{M_q^4+2\lambda\mu^4})$. $\mu$ and $\theta$ are arbitrary integration constants and $M_q$ is defined in Eq.\ (\ref{eq:Mq_0}).
%below \textcolor{red}{Kindly the cite the Eqn. no. where $M_q$ been defined for the reader.}. 
The leading order 2P-correlation function is given, in the same approximation,
\be
\label{eq:G20}
\Box_{dS}G_2^{(0)}(x,x')+3\lambda [G_1^{(0)}(x)]^2 G_2^{(0)}(x,x')+M^2_qG_2^{(0)}(x,x') = -ie^{-3\sqrt{\Lambda}t}\delta^4(x-x').
\ee}

Equation (\ref{eq:G10}) gives the solution of the 1P-function,  thus, by solving Eq.\ (\ref{eq:G20}) for the 2P-function, we are able to completely compute the effective potential.

\subsection{Two-point function}
\label{sec4}

{We first solve Eq.\ (\ref{eq:G20}) and then we analyze the effect on the spectrum arising from the correction in the second order obtained from (\ref{eq:series}). This should grant a more precise determination of the coefficients in the effective potential we aim to determine. We have to solve the equation
\be
\Box_{dS} G_2^{(0)}(x,x')+3\lambda [G_1^{(0)}(x)]^2 G_2^{(0)}(x,x')+M^2_qG_2^{(0)}(x,x') = -ie^{-3\sqrt{\Lambda}t}\delta^4(x-x'),
\ee
where $G_1^{(0)}(x)$ is given by Eq.\ (\ref{eq:G10}) that holds in the rest frame. This implies that we have to solve 
\be
\partial_t^2G_2^{(0)}(x,x')+3\sqrt{\Lambda}\partial_tG_2^{(0)}(x,x')+3\lambda [G_1^{(0)}(t,0)]^2 G_2^{(0)}(x,x')+M^2_qG_2^{(0)}(x,x') = -ie^{-3\sqrt{\Lambda}t'}\delta^4(x-x').
\ee
We introduce the Green's function
\be
\partial_t^2\Delta(t,t')+3\lambda [G_1^{(0)}(t,0)]^2\Delta(t,t')+M^2_q\Delta(t,t') = -ie^{-3\sqrt{\Lambda}t'}\delta(t-t')
\ee
and write the solution of our equation in the form
\be
\label{eq:G200}
G_2^{(0)}(x,x')=\delta^3({\bm x}-{\bm x'})\Delta(t,t')-3\sqrt{\Lambda}\int_{-\infty}^\infty d{\hat t}d^3{\hat x}\sqrt{-g}\delta^3({\bm x}-{\hat{\bm x}})\Delta(t,{\hat t})\partial_{\hat t}G_2^{(0)}({\hat x},x'),
\ee
where
\be
\Delta(t,t')=-i e^{-3\sqrt{\Lambda}t'}\frac{\sqrt{\sqrt{2\lambda\mu^4+M_q^4}+M_q^2}}{\sqrt{2}\sqrt{2\lambda\mu^4+M_q^4}}
\Theta(t-t')\operatorname{cn}(E(t-t')+(4m+1)K(\kappa),\kappa)\operatorname{dn}(E(t-t')+(4m+1)K(\kappa),\kappa),
\ee
being cn and dn Jacobi elliptic functions, $K(\kappa)$ the complete elliptic integral of the first kind, $\Theta(t)$ the Heaviside step function and $m\in\mathbb{Z}$. In momentum space, we obtain (Appendix \ref{appB})
\be\label{boh}
\Delta(\omega,t')=-i e^{-3\sqrt{\Lambda}t'}\frac{E\sqrt{\sqrt{2\lambda\mu^4+M_q^4}+M_q^2}}{\sqrt{2}\sqrt{2\lambda\mu^4+M_q^4}}
\frac{\pi^3}{2K^3(\kappa)\kappa}\sum_{n=0}^\infty(2n+1)^2\frac{q^{n+\frac{1}{2}}}{1-q^{2n+1}}\frac{1}{\omega^2-m_n^2},
\ee
where
\be
\kappa=\sqrt{\frac{M_q^2-\sqrt{M_q^4+2\lambda\mu^4}}{M_q^2+\sqrt{M_q^4+2\lambda\mu^4}}}\,,
\ee
and
\be
q=e^{-\pi\frac{K^*(\kappa)}{K(\kappa)}}
\ee
where $K^*(\kappa)=K(\sqrt{1-\kappa^2})$. The spectrum is given by
\be
\label{eq:m2f}
m_n=(2n+1)\frac{\pi}{2K(\kappa)}E.
\ee
Thus, the Fourier transform of Eq.\ (\ref{eq:G20}) gives
\be
G_2^{(0)}(\omega,{\bm k};x')=e^{i{\bm k}\cdot{\bm x'}}\Delta(\omega,t')-3\sqrt{\Lambda}e^{3\sqrt{\Lambda}t'}\Delta(\omega,t')(-i\omega)G_2^{(0)}(\omega,{\bm k};x')
\ee
that yields the result we are looking for:
\be
\label{eq:G2f}
G_2^{(0)}(\omega,{\bm k};x')=e^{i{\bm k}\cdot{\bm x'}}\Delta(\omega,t')\frac{1}{1+3\sqrt{\Lambda}e^{3\sqrt{\Lambda}t'}\Delta(\omega,t')(-i\omega)}.
\ee
Given the 2P-function we have just obtained, we are now able to evaluate the effective potential.}

%\ag{Need to mention what do we learn from this section before going to next section.}

\medskip

\subsection{Effective potential}
\label{sec5}

{At the leading order, we can evaluate the effective potential in the same way we did in absence of gravity. This yields the following potential:
\be
\label{eq:Veff}
V(\phi)=\frac{1}{2}V_2(\phi-\phi_0)^2-\frac{1}{3!}|V_3|(\phi-\phi_0)^3+\frac{\lambda}{4}\phi^4,
\ee
where $\phi_0=G_1(x)$, $V_2=-\left.G_2^{-1}(p)\right|_{p=0}$ and $V_3=-6\lambda G_1(0)$. For our aims, we move to Euclidean metric and get
\be
\label{eq:V2}
V_2=-i[G_2^{(0)}(0)]^{-1}=-i[\Delta(0)]^{-1}=\frac{\sqrt{2}\sqrt{2\lambda\mu^4+M_q^4}}{\sqrt{\sqrt{2\lambda\mu^4+M_q^4}+M_q^2}}\frac{\pi^2}{4K^2(\kappa)}E A_0\,,
\ee
where $A_0$ is given by
\be
\label{eq:A0}
A_0=\frac{2K^3(\kappa)\kappa}{\pi^3}\left(\sum_{n=0}^\infty\frac{q^{n+\frac{1}{2}}}{1-q^{2n+1}}\right)^{-1}.
\ee
Similarly, one has
\be
\label{eq:V3}
V_3=-6\lambda\sqrt{\frac{2\mu^4}{M_q^2+\sqrt{M_q^4+2\lambda\mu^4}}}\operatorname{sn}\left(\theta,\kappa\right).
\ee
Therefore, we see that the gravity term $\xi/\Lambda$ as expected from Eq.~(\ref{eq:lagrangian}) has a relevant role on the early behavior of the universe as it determines the full shape of the effective potential.}

\subsection{Results and plots}
\label{sec6}

{In order to see where the minima of the potential lie, we observe that $\phi_0\propto\lambda^{-\frac{1}{4}}$. Thus, in the limit $\lambda\rightarrow\infty$, this contribution can be safely neglected. Taking the derivative of eq.(\ref{eq:Veff}), we get
\be
V'(\phi)\approx V_2\phi-\frac{1}{2}|V_3|\phi^2+\lambda\phi^3.
\ee
% and
% \be
% V''(\phi)\approx V_2-|V_3|\phi+3\lambda\phi^2.
% \ee
From the first derivative, we get $\phi=0$ and we have minima when
\be
\frac{1}{4}V_3^2>4\lambda V_2.
\ee
This evaluates to
\be
\operatorname{sn}^2\left(\theta,\kappa\right)>
\frac{4}{9}\frac{1}{2\lambda\mu^4}\sqrt{2}\sqrt{2\lambda\mu^4+M_q^4}\sqrt{\sqrt{2\lambda\mu^4+M_q^4}+M_q^2}\frac{\pi^2}{4K^2(\kappa)}E A_0.
\ee
We introduce the mass gap without gravity as
\be
\label{eq:m0}
m_0^2=\mu^2\sqrt{\frac{\lambda}{2}},
\ee
and we consider $E$ and $m_0$ as independent variables. Indeed, one has
\be
\label{eq:Mq}
M_q^2(E,m_0)=\frac{E^4-m_0^4}{E^2}.
\ee
Therefore, our existence condition just becomes
\be
\label{eq:snth}
\operatorname{sn}^2\left(\theta,\kappa(E,m_0)\right)>
\frac{4}{9}\frac{\sqrt{2}}{4m_0^4}\sqrt{4m_0^4+M_q^4(E,m_0)}\sqrt{\sqrt{4m_0^4+M_q^4(E,m_0)}+M_q^2(E,m_0)}\frac{\pi^2}{4K^2(\kappa(E,m_0))}E A_0(E,m_0),
\ee
where
\be
\kappa(E,m_0)=\sqrt{\frac{M_q^2(E,m_0)-\sqrt{M_q^4(E,m_0)+4m_0^4}}{M_q^2(E,m_0)+\sqrt{M_q^4(E,m_0)+4m_0^2}}}.
\ee
$A_0(E,m_0)$ dependencies are obtained through $\kappa(E,m_0)$ from Eq.\ (\ref{eq:A0}). Therefore, differently from the flat space case, here our range of validity is critically dependent on the product $\xi R$ due to gravity. We can further simplify all these expressions by introducing the ratio $y=M_q/(\sqrt{2}m_0)=\sqrt{(E^4-m_0^4)/(2m_0^2E^2)}$, that is, the excess of the gravitational contribution to the mass with respect to the bare mass gap, where $M_q$ will depend essentially on $\lambda$ and the product $\xi R$. We get
\be
\operatorname{sn}^2\left(\theta,\kappa(y)\right)>
\frac{4}{9}\sqrt{2}\sqrt{1+y^4}\sqrt{\sqrt{1+y^4}+y^2}\frac{\pi^2}{4K^2(\kappa(y))}{\hat E}(y) A_0(y),
\ee
where
\be
\kappa(y)=\sqrt{\frac{y^2-\sqrt{y^4+1}}{y^2+\sqrt{y^4+1}}},
\ee
and
\be
{\hat E}^2(y)=y^2+\frac{1}{2}\frac{1}{y^2+\sqrt{y^4+1}}.
\ee}

\begin{figure}[H]
\centering
\includegraphics{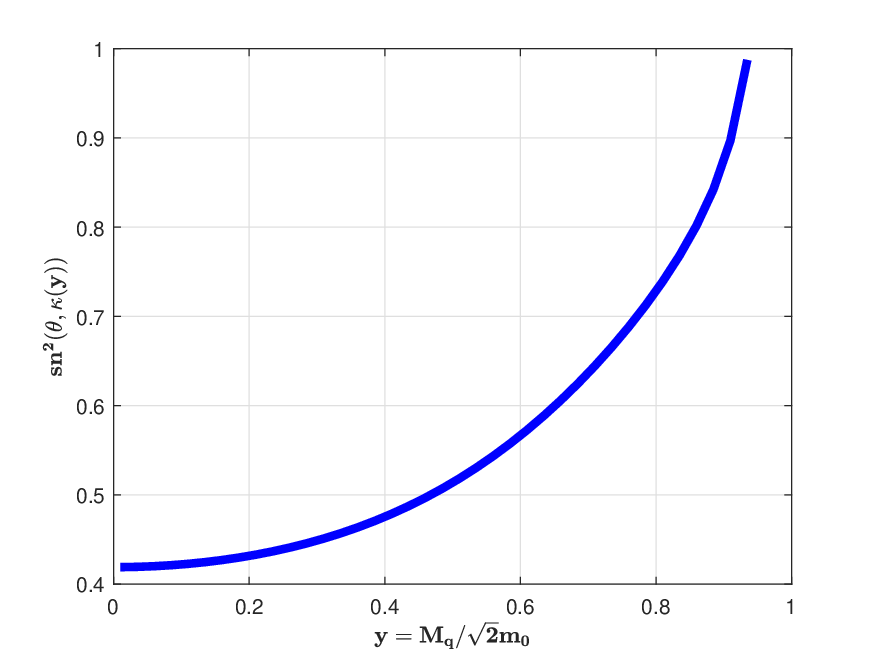}
\caption{\it 
%\ag{Need to format this Fig.} 
Limit on $\operatorname{sn}^2(\theta,\kappa^2(y))$ to obtain minima in the effective potential. \label{fig01}}
\end{figure}

As already pointed out in Ref.~\cite{Czerwinska:2016fky}, assuming a de Sitter background the effect of a non-minimal coupling is to hinder the tunneling effect between the vacua by reducing the depth of the minimum of the false vacuum till washing it out completely if the scalar field is strongly coupled. This should be also compared with \cite{Frasca:2022kfy} in the absence of gravity or with gravity but without non-minimal coupling, already discussed in Sec.~\ref{gravy}. This appears to entail some level of fine tuning with respect to the phase $\theta$ that is not seen in the latter.

\begin{figure}[H]
\centering
\includegraphics{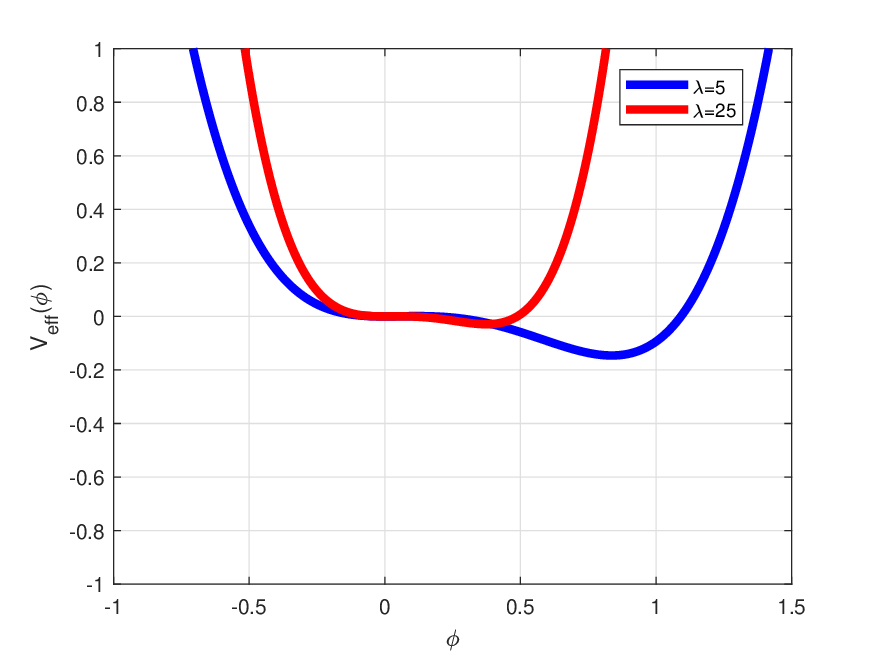}
\caption{\it Effective potential for different values of $\lambda$ within the existence threshold of eq.(\ref{eq:snth}). \label{fig02}}
\end{figure}

\begin{figure}[H]
\centering
\includegraphics{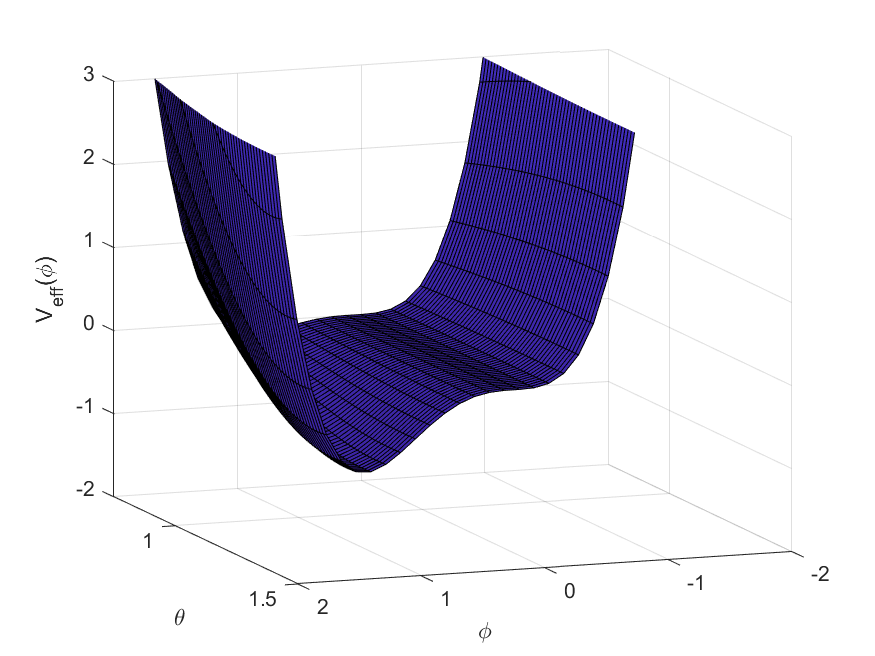}
\caption{\it Effective potential for $\lambda=5$ at $\theta$ varying between the limits of existence of the minima. 
%\ag{range of $\theta$ vaiation ?}
\label{fig03}}
\end{figure}

\begin{figure}[H]
\centering
\includegraphics[width=\textwidth]{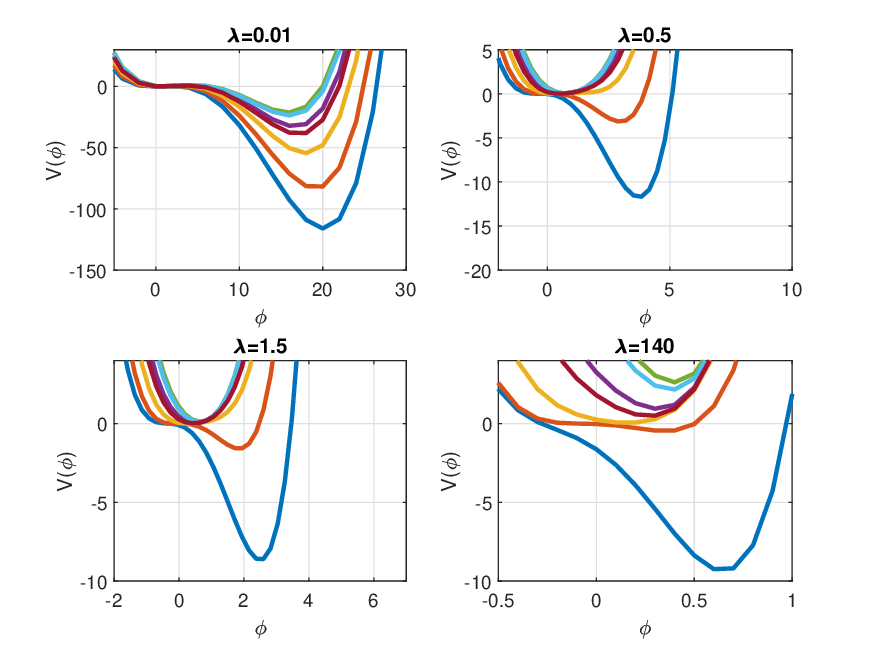}
\caption{\it Plot of $V(\phi)$ as given in eq.~(\ref{eq:Veff}) given $\phi_0$ being the expected vacuum solution. It is seen that the behavior is kept notwithstanding the varying background $\phi_0$ at varying coupling $\lambda$. Different colors represent different curves plotted at various values given by the background $\phi_0$ (see eq.(\ref{eq:G10})) so to give the corresponding spreading. We note that such a spreading decreases with increasing $\lambda$ showing the effect of non-perturbative regimes. 
%\ag{Please mention what color represents which values...} 
Colors represent the potential evaluated at the background solution for the phases $Et+\theta=-0.6021,\ -0.1506,\ 0.3009,\ 0.7523,\ 1.2038,\ 1.6553,\ 2.1068$ respectively.
\label{fig04}}
\end{figure}

\begin{figure}[H]
\centering
\includegraphics[width=\textwidth]{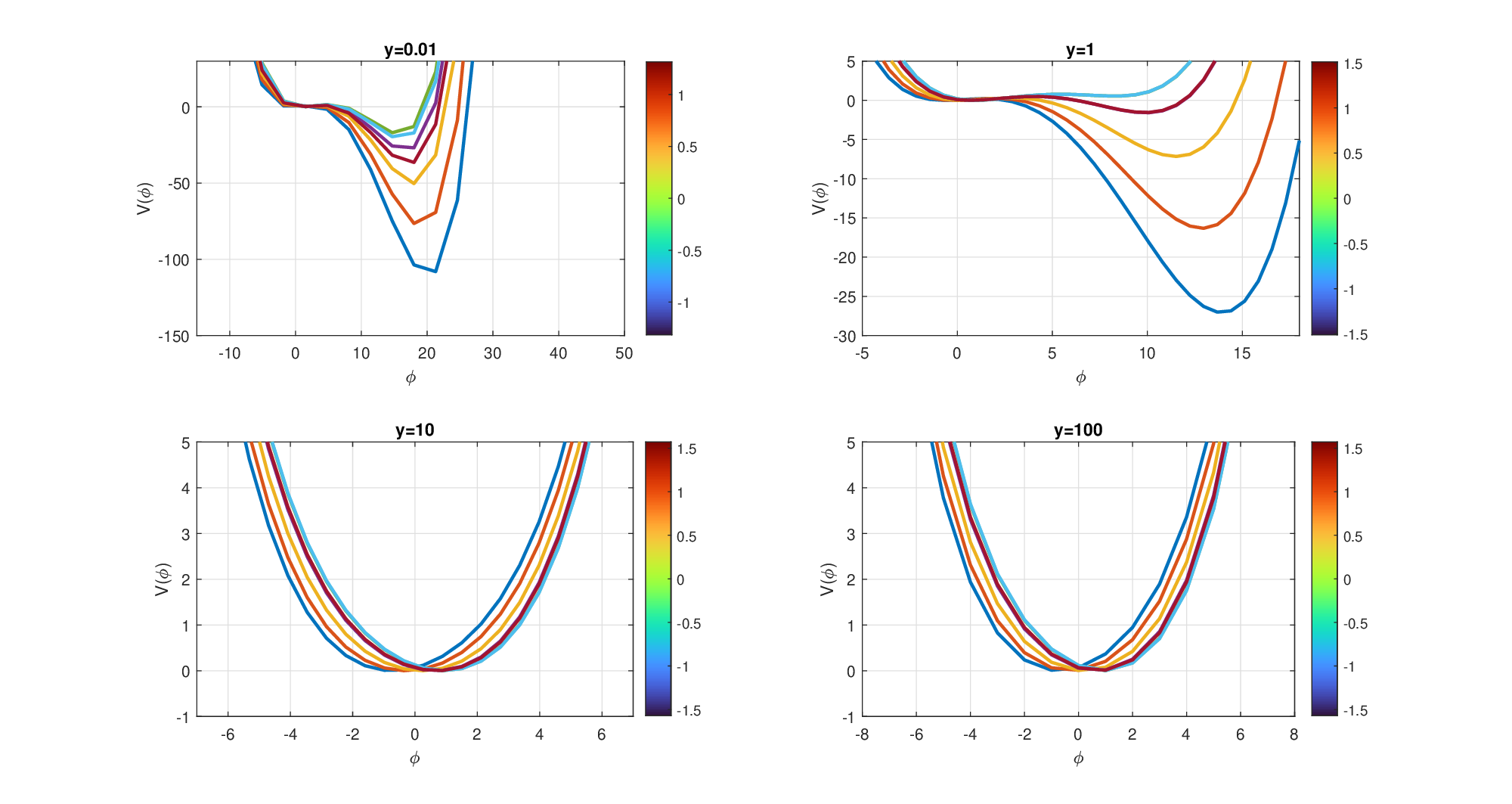}
\caption{\it Same as Fig.~\ref{fig04} but keeping $\lambda$ constant and varying $y$ to evaluate the effect of the $\xi R$ term on the effective potential. It seen that, at increasing relevance of the latter, minima are washed out.
\label{fig05}}
\end{figure}

To understand the relevance of the $\xi R$ term, we should refer to the variable $M_q/(\sqrt{2}m_0)$. Assuming that the shift induced by quantum fluctuations $m_q$ is negligibly small, this boils down to a ratio between $\xi R$ and the mass gap. We keep the latter constant and evaluate the relevance of the $\xi$ parameter. As shown in Fig.~\ref{fig05} by varying the value of $y$ and fixing $\lambda$ to a reasonably small value to get a false vacuum, we can see that the effect of increasing of the term $\xi R$ can wash out the minima.

\medskip

\section{Discussion}\label{disc}

In this paper, we introduced a new technique to 
%compute
evaluate the effective potential for a quartic scalar field theory when a gravitational interactions is present, using exact background solutions. We investigated the false vacuum and its decay. Our main findings are as follows:

\begin{itemize}
    % \item We obtained bounce solutions from effective potentials based on exact solutions and exact Green's functions. Therefore, the results hold for both small and large coupling values (as shown in Eq.\ (\ref{eq:Ueff}) and in Fig.\ \ref{fig3}). 
     \item We obtained bounce solutions from effective potentials using exact solutions and exact Green's functions. Therefore, these results hold both at weak and large couplings (as shown in Eq.\ (\ref{eq:Ueff}) and in Fig.\ \ref{fig3}). 
     \item  We showed the effect of the quartic interaction (see Fig.\ \ref{fig2}) on the false minima. As the interaction strength increases, it is easier to tunnel and the vacuum decays quickly. We also showed the effect of the initial $\theta$ (see Fig.\ \ref{fig3}) on the false minima for fixed $\lambda$. The choice of initial $\theta$ does not alter the probability of false-vacuum decay.
    \item We compared the vacuum decay of the strongly coupled theory to that of Coleman's bounce in Eqs.\ (\ref{eq:comp1}) and (\ref{eq:comp2}). In the limit when the interaction strength is weak, for small values of $\lambda$ the solutions become identical.
    \item We showed that 
    %in the strong coupling regime, the quartic interaction 
    a strongly coupled quartic interaction
    leads to a decay of the false vacuum also in the presence of gravitational interactions. 
    The evaluation of the decay rate was accomplished using the exact Green's function in the non-perturbative regime
    %The decay rate was evaluated by following our exact Green's function in the non-perturbative regime, 
    as given in Eqs.\ (\ref{eq:B1}) and (\ref{eq:B21}). 
%    \item We showed how false vacuum is developed in the strongly coupled regime (see Fig. \ref{fig1}).
     \item We compared the vacuum decay rate with and without the inclusion of gravity effects in Eqs.\ (\ref{comp:gr1}) and (\ref{comp:gr2}). The effect of Einstein gravity is unidirectional inasmuch as it always facilitates the decay from the false vacuum, regardless of whether the latter has positive or negative energy. This may not be true in other gravitational theories such as Starobinsky's, ${\cal L}\propto \mathcal{R}+\alpha \mathcal{R}^2$ \cite{Vicentini:2020lhm}. It might be worth generalizing these results to $f(R)$ or other, more realistic models including the Ricci tensor in the action.
     \item In the $\Lambda \rightarrow \infty$ limit, the gravitational interactions become negligible and the solution becomes the same as that without gravity.
     \item The running of the coupling of the self-interaction of the scalar field is obtained by comparing the strongly and weakly coupled regimes. This has the  behavior already shown in \cite{Frasca:2013tma} %and shows that the theory becomes weakly coupled in the infrared limit
     with the theory becoming weakly coupled at lower energies (infrared limit). 
     %This proves that Einstein gravity made the weak coupling approximation even more reliable as the universe cooled down. 
     This shows that the effect of Einstein gravity is to render the weak-coupling approximation more reliable when the universe cools down.
     Besides, the origin of this scalar sector could be traced back to a conformally flat limit of the Einstein equations.
% Added aftere referee's comments
     \item 
     We also analyzed the case of a non-minimal coupling between gravity and the scalar field. The general effect is to hinder the tunneling between the vacua. If the scalar field is strongly coupled, that is large $\lambda$, the effect can be completely washed out (see Fig.~\ref{fig4}). Similarly, by keeping $\lambda$ small and constant and varying the term $\xi \phi ^2 \mathcal{R}$, we get an even more marked effect of washing out of the minima (see Fig.~\ref{fig5}).
     %} \textcolor{red}{Just like the other points let us highlight the most important expressions and the plots we got for this point.}
%
\end{itemize}
  Thanks to the exact Green's function technique applied to the effective potential, the study of the fate of false vacuum is now free from the limitations of perturbation theory and its range of applications can be enlarged to non-perturbative regimes in quantum field theory. The method developed so far is applicable to ultracold atoms and Bose gas \cite{Billam:2021nbc} and quantum spin chain systems \cite{Lagnese:2021grb}, as well as to other condensed matter systems involving strong interactions in general. In particular, we can extend the analysis to non-perturbatve strongly-coupled beyond-the-Standard-Model theories in the presence of gravitational interactions.  We envisage this to have a remarkable impact in primordial cosmological scenarios with first-order phase transitions, %such as old inflation or, more importantly, 
    especially for the search of a stochastic gravitational-wave background from such a primordial source. The LIGO--Virgo--KAGRA network has already constrained this scenario \cite{Romero:2021kby,LIGO-result}, but the result is model-dependent and this new non-perturbative window could alter the prediction of the model. A detailed analysis of phase transitions and gravitational-wave predictions following our non-perturbative method is beyond the scope of this paper and should be taken up in the future.

 \section*{Acknowledgement}

AG acknowdedges useful feedback from Marek Lewicki. We would like to thank Marek Lewicki, Alberto Salvio and Alessandro Strumia for very useful comments.

\medskip

\appendix

\section{Zero mode}\label{appA}

%The Hamiltonian of the system is given by
We consider the following Hamiltonian
\begin{equation}
    H=\int d^3x\left[\frac{1}{2}(\partial_t\phi)^2+\frac{1}{2}(\nabla\phi)^2+\frac{\lambda}{4}\phi^4\right],
\end{equation} 
%We 
and linearize it around the classical solution (see
%given by 
$G_1$ in (\ref{solG1}))
\begin{equation}
    \phi(x)=\phi_0(x)+\delta\phi(x)\,,
\end{equation}
yielding
\begin{equation}
   H=H_0+\int d^3x\left[\frac{1}{2}(\partial_t\delta\phi)^2+\frac{1}{2}(\nabla\delta\phi)^2
   +\frac{3}{2}\lambda\phi_0^2\delta\phi^2\right]+O\left(\delta\phi^3\right).
\end{equation}
%being 
$H_0$ 
%the contribution coming from
is due to
the classical solution. 
%The linear part can be diagonalized with a Fourier series provided we are able to get the eigenvalues and the eigenvectors of the operator
We are left with a linear term that we can diagonalize using a Fourier series if we know the eigenvalues and the eigenvectors of the operator
\begin{equation}
   L_{\mu_0^2=0}=-\Box+3\lambda\phi_0^2(x).
\end{equation}
% It is not difficult to realize that there is a zero mode. We give the solutions for both the zero and non-zero modes. The spectrum is continuous with eigenvalues 0 and $3\mu^2\sqrt{\lambda/2}$ with $\mu$ varying continuously from 0 to infinity. The zero-mode solution has the aspect
We see that there is a zero mode. Indeed, we can see that the spectrum is given by 0 and $3\mu^2\sqrt{\lambda/2}$ where $\mu$ is a constant running continuously from 0 to $\infty$. The zero mode is given by the formula
\begin{equation}
  \chi_0(x,\mu)=a_0\,{\rm cn}(p\cdot x+\theta,i)\,{\rm dn}(p\cdot x+\theta,i)
\end{equation}
%being 
where
$a_0$ 
is
a normalization constant. 
%Non-zero modes are given by
Similarly, for non-zero modes we get
\begin{equation}
  \chi(x,\mu)=a'\,{\rm sn}(p\cdot x+\theta,i)\,{\rm dn}(p\cdot x+\theta,i).
\end{equation}
%with 
where
$a'$ 
%again 
is
a normalization constant. These hold on-shell
%, that is when 
for
$p^2=\mu^2\sqrt{\lambda/2}$. 
Since the spectrum is continuous, these eigenfunctions are not normalizable. Therefore, we note that there is a doubly degenerate set of zero modes spontaneously breaking translational invariance and the $Z_2$ symmetry of the theory. 
%As seen in Sec.\ref{sec:mass}, the phases $\theta$ are fixed to the values $(4n+1)K(i)$ in the massless case. 
This gives for the zero mode
\begin{equation}
  \chi_0(x,\mu)=-2a_0\frac{{\rm sn}(p\cdot x,i)}{{\rm dn}^2(p\cdot x,i)}.
\end{equation}
%For a given $\mu$ parameter, $Z_2$ symmetry is spontaneously broken through this zero mode. This mode disappears when $\mu=0$, as it should, and one goes back to a standard textbook solution.

\section{Calculation of (\ref{boh})}\label{appB}

We want to evaluate the following Green's function in momentum space:
\be
\Delta(t,t')=-i e^{-3\sqrt{\Lambda}t'}\frac{\sqrt{\sqrt{2\lambda\mu^4+M_q^4}+M_q^2}}{\sqrt{2}\sqrt{2\lambda\mu^4+M_q^4}}
\Theta(t-t')\operatorname{cn}(E(t-t')+(4m+1)K(\kappa),\kappa)\operatorname{dn}(E(t-t')+(4m+1)K(\kappa),\kappa).
\ee
Firstly, we observe that
\be
\operatorname{cn}(z,\kappa)\operatorname{dn}(z,\kappa)=\operatorname{sn}'(z,\kappa),
\ee
where the derivative is respect to $z$. Then,
\be
\operatorname{sn}(z,\kappa)=\frac{2\pi}{K(\kappa)\kappa}\sum_{n=0}^\infty\frac{q^{n+\frac{1}{2}}}{1-q^{2n+1}}\sin\left((2n+1)\frac{\pi}{2K(\kappa)}z\right),
\ee
where
\be
q=e^{-\pi\frac{K^*(\kappa)}{K(\kappa)}}
\ee
where $K^*(\kappa)=K(\sqrt{1-\kappa^2})$. Therefore, we have straightforwardly
\bea
&&\Delta(t,t')=-i e^{-3\sqrt{\Lambda}t'}\frac{\sqrt{\sqrt{2\lambda\mu^4+M_q^4}+M_q^2}}{\sqrt{2}\sqrt{2\lambda\mu^4+M_q^4}}
\Theta(t-t')
\frac{\pi^2}{K^2(\kappa)\kappa}\sum_{n=0}^\infty(2n+1)\frac{q^{n+\frac{1}{2}}}{1-q^{2n+1}}\times \nonumber \\
&&\cos\left((2n+1)\frac{\pi}{2K(\kappa)}(E(t-t')+(4m+1)K(\kappa))\right)= \nonumber \\
&&i e^{-3\sqrt{\Lambda}t'}\frac{\sqrt{\sqrt{2\lambda\mu^4+M_q^4}+M_q^2}}{\sqrt{2}\sqrt{2\lambda\mu^4+M_q^4}}
\Theta(t-t')
\frac{\pi^2}{K^2(\kappa)\kappa}\sum_{n=0}^\infty(-1)^n(2n+1)\frac{q^{n+\frac{1}{2}}}{1-q^{2n+1}}\times \nonumber \\
&&\sin\left((2n+1)\frac{\pi}{2K(\kappa)}E(t-t')\right).
\eea
A Fourier transform in time will yield
\be
\Delta(\omega,t')=-i e^{-3\sqrt{\Lambda}t'}\frac{E\sqrt{\sqrt{2\lambda\mu^4+M_q^4}+M_q^2}}{\sqrt{2}\sqrt{2\lambda\mu^4+M_q^4}}
\frac{\pi^3}{2K^3(\kappa)\kappa}\sum_{n=0}^\infty(2n+1)^2\frac{q^{n+\frac{1}{2}}}{1-q^{2n+1}}\frac{1}{\omega^2-m_n^2}.
\ee
where we have set $m_n=(2n+1)\pi E/2K(\kappa)$.

% \bea
% &&\frac{\sqrt{\sqrt{2\lambda\mu^4+M_q^4}+M_q^2}}{\sqrt{2}\sqrt{2\lambda\mu^4+M_q^4}}
% \sqrt{\frac{M_q^2+\sqrt{M_q^4+2\lambda\mu^4}}{M_q^2-\sqrt{M_q^4+2\lambda\mu^4}}}
% \sqrt{M_q^2+\frac{\lambda\mu^4}{M_q^2+\sqrt{M_q^4+2\lambda\mu^4}}}= \nonumber \\
% &&\frac{\sqrt{\sqrt{2\lambda\mu^4+M_q^4}+M_q^2}}{\sqrt{2}\sqrt{2\lambda\mu^4+M_q^4}}
% \sqrt{\frac{M_q^2+\sqrt{M_q^4+2\lambda\mu^4}}{M_q^2-\sqrt{M_q^4+2\lambda\mu^4}}}
% \sqrt{\frac{M_q^4+M_q^2\sqrt{M_q^4+2\lambda\mu^4}+\lambda\mu^4}{M_q^2+\sqrt{M_q^4+2\lambda\mu^4}}}= \nonumber \\
% &&\frac{1}{\sqrt{2}\sqrt{2\lambda\mu^4+M_q^4}}
% \sqrt{\frac{M_q^2+\sqrt{M_q^4+2\lambda\mu^4}}{M_q^2-\sqrt{M_q^4+2\lambda\mu^4}}}
% \sqrt{M_q^4+M_q^2\sqrt{M_q^4+2\lambda\mu^4}+\lambda\mu^4} \nonumber \\
% \eea

\end{document}